\documentclass[aps, prb, reprint, floatfix, a4paper, amsmath, amssymb, showpacs]{revtex4-1} 
\usepackage{graphicx}
\usepackage{multirow}
\usepackage{textcomp} 
\PassOptionsToPackage{caption=false}{subfig}
\usepackage{subfig}
\usepackage{notoccite}

\newcommand{\LB}{4.36}
\newcommand{\LBerror}{ $\pm0.05$}
\newcommand{\HB}{5.97}
\newcommand{\HBerror}{ $\pm0.01$}
\newcommand{\Lphase}{$-0.48 \pm0.06$}
\newcommand{\Hphase}{$-0.20 \pm0.05$}
\newcommand{\SdHphase}{$-0.48 \pm0.01$}
\newcommand{\LphaseFFT}{$-0.45 \pm0.25$}
\newcommand{\HphaseFFT}{$-0.25 \pm0.25$}
\newcommand{\SdHphaseFFT}{$-0.49 \pm0.25$}

\newcommand{\figwidth}{0.45\textwidth}

\begin{document}

\title{Millikelvin de Haas--van Alphen and Magnetotransport studies of Graphite}

\author{S.B. Hubbard}\author{T.J. Kershaw}\author{A. Usher}\author{A.K. Savchenko}\author{A. Shytov}

\affiliation{School of Physics, University of Exeter, Stocker Road, Exeter EX4 4QL}

\date{\today}

\begin{abstract}
Recent studies of the electronic properties of graphite have produced conflicting results regarding the positions of the different carrier types within the Brillouin zone, and the possible presence of Dirac fermions.  In this paper we report a comprehensive study of the de Haas--van Alphen, Shubnikov--de Haas and Hall effects in a sample of highly orientated pyrolytic graphite, at temperatures in the range 30$\,$mK to 4$\,$K and magnetic fields up to 12$\,$T. The transport measurements confirm the Brillouin-zone locations of the different carrier types assigned by \citet{Schroeder1968}: electrons are at the $K$-point, and holes are near the $H$-points. We extract the cyclotron mass and scattering time for both carrier types from the temperature- and magnetic-field-dependences of the magneto-oscillations.  Our results indicate that the holes experience stronger scattering and hence have a lower mobility than the electrons.  We utilise phase--frequency analysis and intercept analysis of the $1/B$ positions of magneto-oscillation extrema to identify the nature of the carriers in graphite, whether they are Dirac or normal (Schr\"odinger) fermions. These analyses indicate normal holes and electrons of indeterminate nature.
\end{abstract}


\maketitle

\section{Introduction}

Graphite is a fascinating material whose novel electronic properties have been extensively studied.\cite{Brandt1988}  It consists of weakly-bonded layers of graphene, resulting in a highly anisotropic Fermi surface and semi-metallic properties.  Recently there has been a resurgence of interest in graphite due to
the possible occurrence of quasi-relativistic graphene physics in this
3D bulk material. There is however much discussion over whether
the Dirac fermions found in graphene are actually present in graphite
samples.\cite{Lukyanchuk2004,Lukyanchuk2006,Mikitik2006,Orlita2008,Schneider2009,Lukyanchuk2010,Schneider2010}

Tight-binding calculations using a 2D model of graphite were originally utilised by \citet{Slonczewski1958} (SW) in 1958. They introduced the first 3D model of graphite, a $\mathbf{k}\cdot\mathbf{p}$ perturbation calculation using the tight-binding wavefunctions of the 2D model as basis functions. The result was a band structure dependent on seven parameters ($\gamma_{0}$ to $\gamma_{5}$ and $\Delta$) to be determined experimentally. 

\begin{figure}
\includegraphics[width=\figwidth]{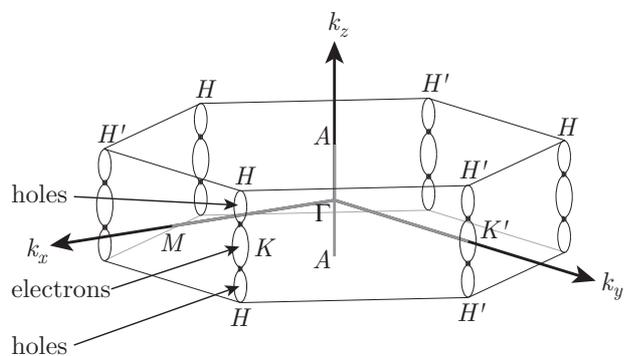}

\caption{Brillouin zone of graphite. Electron and hole pockets formed by
the $\pi$ bands are centred along the zone edges $HKH$.}
\label{Fig:brillouin_zone}
\end{figure}

Table~\ref{Flo:previous_conclusions_table} summarises the results of subsequent studies of the band structure of graphite, which utilise magneto-oscillatory effects such as the de Haas--van Alphen (dHvA) and Shubnikov--de Haas (SdH) effects. \citet{McClure1957} used an interpretation of previous measurements of the dHvA effect in single-crystal natural graphite\cite{Shoenberg1952,Berlincourt1955} to determine, or place limits on, the SW parameters. He derived a Fermi surface of the form of three approximately ellipsoidal surfaces along the $HKH$ edge of the Brillouin zone.  This analysis was unable to specify the parameter $\gamma_{2}$, determining the carrier types (electrons or holes) at the three extremal orbits, but favoured the assignment of holes to the extremal orbit at $K$ and electrons to the extremal orbits close to $H$. Figure~\ref{Fig:brillouin_zone} shows the Brillouin zone of graphite with the ellipsoidal Fermi surfaces along the $HKH$ edges. Note that in the figure the carrier labelling reflects that of the current accepted associations from the work of \citet{Schroeder1968}, which contradict the original assignments of \citeauthor{McClure1957}.

Subsequently, \citet{Soule1964} used angle-dependent SdH measurements on single-crystal natural graphite to demonstrate that the Fermi surfaces for both holes and electrons are closed. They found that the extremal cyclotron orbit at the $K$ point is larger than the orbit near the $H$ point.

\citet{Williamson1965} were the first authors to consider the phase of the magneto-oscillations.  They followed the  formula of \citet{Lifshitz1956} from which the fundamental harmonic of the oscillatory magnetic susceptibility can be written as

\squeezetable
\begin{table*}
\begin{centering}
\setlength{\tabcolsep}{2ex}
\begin{ruledtabular}
\begin{tabular}{c c c ccc ccc}

First 							& Material 	& Technique 			& \multicolumn{3}{c}{Low frequency} 					& \multicolumn{3}{c}{High frequency} 					\\
Author						& studied		& 	used				& $f_{\mathrm{FFT}}$ 	
																		& Carrier, 		& Total phase, 				& $f_{\mathrm{FFT}}$ 	
																														& Carrier, 		& Total phase,				\\
							& 			& 					& (T) 		& location 	& $\varphi=\left(\delta-\gamma\right)\,\left[2\pi\right]$
																							 				& (T) 		& location 	& $\varphi=\left(\delta-\gamma\right)\,\left[2\pi\right]$\\
							& 			& 					&			& 			& nature					& 			& 		 	&nature					\\
\hline\\[-5pt]
McClure\protect\cite{McClure1957} 		& -	 		& dHvA Theory 		& -		 	& electrons	& - 						& -			& holes		&-						\\
							& 			& 					&			& near $H$	& 						& 			& $K$	 	&						\\[1ex]

Soule\protect\cite{Soule1964} 			& Natural 		& Angle-  				& 4.8	 	& electrons 	& - 						& 6.7		& holes		&-						\\
							& 			& dependent SdH		&			& near $H$	& 						& 			& $K$	 	&						\\[1ex]

Williamson\protect\cite{Williamson1965} 	& Pyrolytic	& dHvA 				& 4.8	 	& electrons 	& $0.38 \pm0.05$ 			& 6.6		& holes		& $0.32 \pm0.09$			\\
							&			& 					& $\pm0.3$	& near $H$	& 3D SF					& $\pm0.4$	& $K$	 	& 3D SF					\\[1ex]

Schroeder\protect\cite{Schroeder1968} 	& Pyrolytic	& Polarised	 		& -	 		& holes 		& - 						& -			& electrons	&-						\\
							& 			& magnetoreflection		&			& near $H$	& 						& 			& $K$	 	&						\\[1ex]

Woollam\protect\cite{Woollam1971}		& Pyrolytic	& SdH, Hall effect, 		& 4.9 		& holes 		& -			 			& 6.2		&electrons	&-						\\
 							& 			& thermopower and 		&  $\pm0.1$	& near $H$ 	&  						& $\pm0.3$	&$K$		&						\\
 							& 			& thermal resistivity 		&  			&  			&  						& 			&			&						\\[1ex]
									
Luk'yanchuk\protect\cite{Lukyanchuk2004} & HOPG 		& dHvA and SdH 		& 4.68 		& electrons 	& 0.375 					& 6.41		& holes		& 0.5					\\
							& 			& 					&			&			& 3D SF					& 			& 		 	& 2D DF$^\ast$			\\[1ex]
									
Luk'yanchuk\protect\cite{Lukyanchuk2006} & HOPG		& SdH 				& 4.68 		& electrons 	& 0.5					& 6.41		& holes		& 0						\\
							& 			& 					&			& 			& 3D SF$^\ast$			& 			& 		 	& 2D DF					\\[1ex]
									
Mikitik\protect\cite{Mikitik2006} 			& -	 		& Re-analysis of 
									[\protect\onlinecite{Lukyanchuk2004}] 	& -	 		& holes	 	&  0.375					& -			& electrons	&0.5						\\
							& 			& 					&			& near $H$	& 3D SF					& 			& $K$		& 3D SF$^\ast$			\\[1ex]
									
Orlita\protect\cite{Orlita2008} 			& HOPG		& Magnetotransmission 	& -			& holes	 	& -	 					& -			& electrons	& -						\\
							& 			& 					&			& at $H$		& 3D DF					& 			& $K$	 	& SF						\\[1ex]
									
\multirow{3}{*}{Schneider\protect\cite{Schneider2009}} 		
							& Natural		& SdH 				& 4.51 		& holes	 	& $-0.43 \pm0.05$			& 6.14		& electrons	& $-0.28 \pm0.05$			\\
							& and		& 					& $\pm0.05$	& near $H$	& 3D SF$^\ast$			& $\pm0.05$	& $K$	 	& 3D SF$^\ast$			\\[1ex]
							& HOPG		& 					& 			& 			& $-0.52 \pm0.05$			& 			& 		 	&$-0.46 \pm0.05$ 			\\
							& 			& 					& 			& 			& 2D SF					& 			& 		 	& 2D SF					\\[1ex]
									
\bf{This paper:}					& HOPG 		& dHvA, Hall and SdH  	& \LB 		& holes 		& \Lphase					& \HB 		& electrons 	& \Hphase	 			\\
 							& 			& 				 	& \LBerror		& near $H$	& 2D SF 					& \HBerror	& $K$ 		& indeterminate 			\\[1ex]
							
 \end{tabular}
\end{ruledtabular}
\par\end{centering}
\caption{Comparison of the conclusions of studies of the carriers' locations within the Brillouin zone and their natures in graphite. $f_{\mathrm{FFT}}$ is the frequency of the $1/B$ oscillations for each carrier type.  The phases $\delta$ and $\gamma$ are defined in the text.  In the case of Williamson, the authors did not discuss carrier natures, but their phases are consistent with the interpretations stated.  In some other cases authors' conclusions are not consistent with the phases measured.  These are marked ($^\ast$) and discussed in the text.}
\label{Flo:previous_conclusions_table}
\end{table*}
\vspace{-0.5 cm}
\begin{equation}
\Delta\chi\propto\cos\left(2\pi\left[\frac{B_{0}}{B}-\gamma+\delta\right]\right),
\label{eq:delta_chi_form}
\end{equation}
where $B_0$ is the fundamental frequency of the magneto-oscillations for a given carrier type, and $B$ is the applied magnetic field. The phase factor $\gamma$ comes from the Onsager--Lifshitz quantization condition and is $\frac{1}{2}$ for non-relativistic free electrons. The offset $\delta$ is related to the curvature of the Fermi surface in the $k_z$-direction, and is $-\frac{1}{8}$ or $+\frac{1}{8}$ for maximum or minimum extremal cross-sections, respectively. \citeauthor{Williamson1965} showed that, according to the SW model, $\gamma$ takes on its usual value of $\frac{1}{2}$ at the $K$ point of the Brillouin zone, but is zero at the $H$ points.  However, at the positions of both the extremal orbits ($K$, and $\sim$70\% of the way from $K$ to $H$)\cite{McClure1957} the SW model predicts $\gamma=\frac{1}{2}$ -- a result with which their experimental dHvA results agreed.  \citeauthor{Williamson1965} also pointed out a fundamental difficulty in measuring the phase: one needs to approach the quantum limit of low Landau-level filling factors in order to obtain accurate results, but in this limit the oscillations cease to be periodic in $1/B$.

In 1968, the magnetoreflection studies of \citet{Schroeder1968} led to a reassessment of the SW band parameters, with the conclusion that the majority carrier types assigned previously were incorrect. It was shown that \textit{electrons} occupy the orbits at the $K$ point while \textit{holes} occupy those near the $H$ point (as shown in Fig.~\ref{Fig:brillouin_zone}). This was confirmed by the magnetoresistance, Hall effect, thermopower and thermal resistivity measurements of \citeauthor{Woollam1971}.\cite{Woollam1971}

Since the fabrication of individual graphene layers \cite{Novoselov2004} the experimental interest in graphene has increased tremendously. Experiments have shown that the charge carriers in graphene are massless, quasi-relativistic Dirac fermions (DFs) with a linear dispersion relation resulting in an anomalous quantum Hall effect with plateaus at half-integer filling factors.\cite{Novoselov2005,Zhang2005} Recently \citet{Lukyanchuk2004,Lukyanchuk2006} have reported 2D-like electronic properties in highly oriented pyrolitic graphite (HOPG) and have claimed also to have observed the presence of DFs. In 2004 \citet{Lukyanchuk2004} presented dHvA and SdH experiments on a sample of HOPG at a temperature of 2$\,$K using magnetic fields up to 9$\,$T. They determined the nature of the carriers, either massless DFs or massive Schr\"odinger fermions (SFs, also described in the literature as ``normal'' fermions), by two-dimensional phase-frequency analysis of the complex Fourier transforms of the quantum oscillations. Table~\ref{Flo:carrier_nature_table} details the values of the phases $\gamma$ and $\delta$ from Eq. \ref{eq:delta_chi_form} expected for carriers of different nature (DF or SF) and dimensionality (2D or 3D).  They concluded that the high-frequency carriers (which they assigned to be holes) are 2D DFs and the low-frequency carriers (which they assigned to be electrons) are 3D SFs.  In this terminology the conclusion of the previous studies of \citet{Williamson1965} were that both carrier types were 3D SF, raising a controversy over the nature of the high-frequency carriers. \citeauthor{Lukyanchuk2004} assigned the carrier types in contradiction to \citeauthor{Schroeder1968} and \citeauthor{Woollam1971}. 

\begin{table}
\begin{centering}
\setlength{\tabcolsep}{2ex}
\begin{ruledtabular}
\begin{tabular}{c c c c}
Nature & $\gamma\,\left[2\pi\right]$ & $\delta\,\left[2\pi\right]$ & Total phase $\varphi=\left(\delta-\gamma\right)\,\left[2\pi\right]$\\[2ex]
\hline\\[-5pt]
3D SF	& $\pm0.5$ 	& -0.125 	& +0.375, -0.625\\
2D SF	& $\pm0.5$ 	& 0 		& +0.5, -0.5\\
3D DF	& 0 			& -0.125 	& +0.875, -0.125\\
2D DF	& 0 			& 0 		& 0\\
\end{tabular}
\end{ruledtabular}
\par\end{centering}
\caption{Phases $\gamma$ and $\delta$ from Eq. \eqref{eq:delta_chi_form} for different carrier natures: Schr\"odinger or Dirac fermions (SF or DF); two- or three-dimensional (2D or 3D). }
\label{Flo:carrier_nature_table}
\end{table}

\begin{figure}
\subfloat[]{\includegraphics[width=0.435\textwidth]{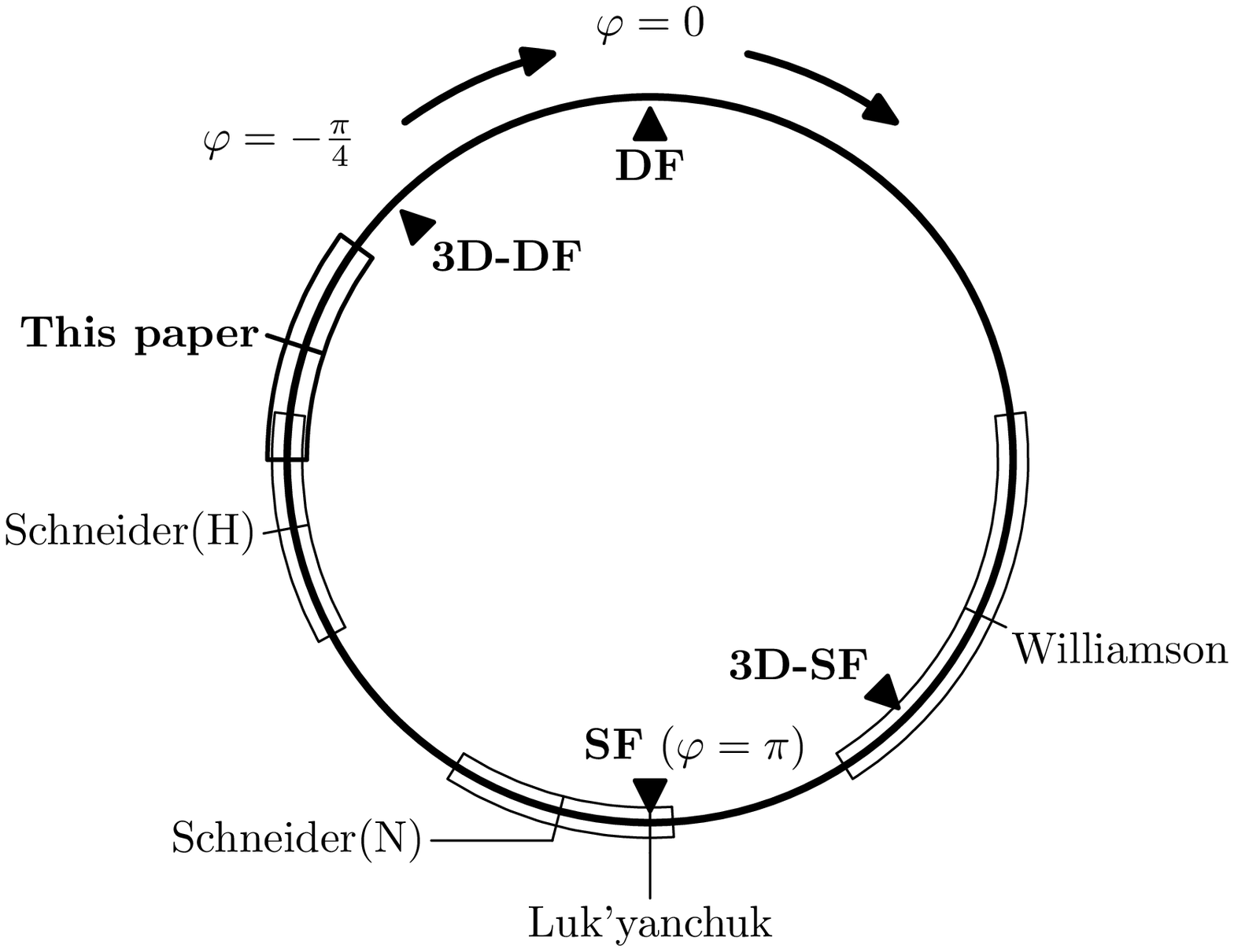}
\hspace{7.5pt}
\label{Flo:phase_summary_HF}}\\
\subfloat[]{\hspace{10pt}\includegraphics[width=0.396\textwidth]{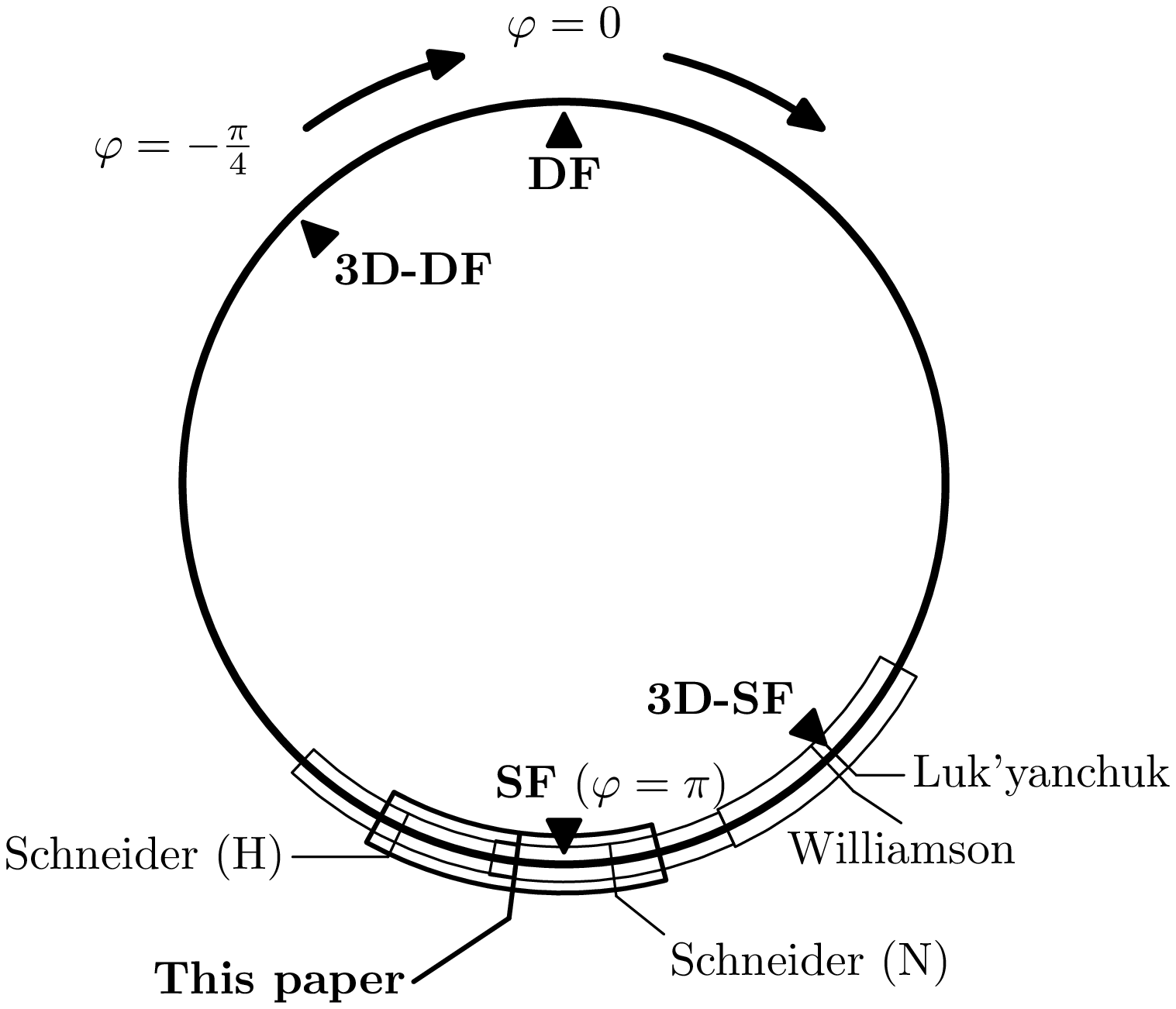}
\label{Flo:phase_summary_LF}}\\

\caption{Graphical representation of the total phases $\varphi=\left(\delta-\gamma\right)\,\left[2\pi\right]$ determined by the authors listed in Table~\ref{Flo:previous_conclusions_table}, (a) for high-frequency carriers, and (b) for low-frequency carriers. The phase starts at zero at the top of each circle and increases in the clockwise direction.  In the case of the measurements of \citeauthor{Schneider2009}, (N) refers to their results on natural graphite and (H) refers to their results on HOPG.}
\label{Fig:phase_summary}
\end{figure}

In 2006, \citet{Lukyanchuk2006} detailed further SdH experiments under similar conditions. The carriers were again assigned in contradiction to \citeauthor{Schroeder1968} and \citeauthor{Woollam1971}. Plotting the extrema of the oscillations against Landau-level index, the total phase $\left(\delta-\gamma\right)$ for each carrier type was found using the equation 
\begin{equation}
\Delta\sigma_{xx}\left(B\right)\simeq A\left(B\right)\cos\left(2\pi\left[\frac{B_{0}}{B}-\gamma+\delta\right]\right),
\label{eq:delta_sigma_form}
\end{equation}
where $\Delta\sigma_{xx}$ is the oscillatory part of the longitudinal conductivity and $A\left(B\right)$ is the non-oscillatory amplitude.  Hence the carrier nature and dimensionality was deduced from Table~\ref{Flo:carrier_nature_table}. The conclusion was in agreement with the authors' previous letter:\cite{Lukyanchuk2004} they found that the high-frequency carriers (which were assigned to be holes) are 2D DFs and the low-frequency carriers (which were assigned to be electrons) are 3D SFs. This intercept analysis method can also be applied to dHvA oscillations through Eq. \eqref{eq:delta_chi_form}.

\citet{Mikitik2006} identified an incorrect assertion by \citeauthor{Lukyanchuk2004} that there is an inherent phase shift of $\pi$ in the magnetic susceptibility for holes compared to that of electrons. They re-analysed the data, concluding that both carriers are 3D SFs.  However, the phase for the high-frequency carriers (electrons), $0.5\,\left[2\pi\right]$ would be more consistent with 2D SFs.   They also related the value of $\gamma$ to the number of band-contact lines encircled by the carrier orbits in $k$-space: $\gamma=\frac{1}{2}$ or $0$ when an even or odd number of band-contact lines are encircled, respectively. Both extremal orbits in graphite enclose four band-contact lines and so one should expect $\gamma=\frac{1}{2}$ (the carriers should be SFs).

\citet{Orlita2008} performed optical magnetotransmission experiments on HOPG.  These measurements are sensitive to the carriers at the $H$ and $K$ points, not at the extremal cross-sections probed by dHvA and SdH.  They confirmed that the carriers at the $H$ point (assumed to be holes)  had a $\sqrt{B}$ Landau-level spectrum and are therefore DFs, with some indication of a 3D nature. A linear Landau-level spectrum was also observed and assigned to electrons at $K$.

In 2009, \citet{Schneider2009} performed SdH experiments on both natural graphite and HOPG at millikelvin temperatures. They utilised the phase--frequency analysis technique to determine carrier nature. In HOPG their results were consistent with the conclusion that both carriers are 2D SFs. In natural graphite, they concluded that both holes and electrons are 3D SFs. However, the total phase, $\left(\delta-\gamma\right)$ for the electron oscillations was $-0.28\,\left[2\pi\right]$; $\delta=-\frac{\pi}{4}$, and so $\gamma\simeq-0.16\,\left[2\pi\right]$ which is not consistent with either carrier nature, though it is closer to DF. Furthermore, the total phase found for the holes was $-0.43\,\left[2\pi\right]$; with $\delta=-\frac{\pi}{4}$ this gives $\gamma\simeq-0.31$ which is also not consistent with either carrier nature. \citeauthor{Schneider2009} once again raised the issue of whether an intercept analysis of the type used by \citet{Lukyanchuk2006} can yield an accurate value of the phase.  This remains a hotly contested issue.\cite{Lukyanchuk2010,Schneider2010}

Figure~\ref{Fig:phase_summary} is a graphical representation of the total phases $\varphi$ reported for the carriers in graphite by the authors listed in Table~\ref{Flo:previous_conclusions_table}.  The phases for 2D DFs and SFs are shown at the tops and bottoms of the phase circles, respectively.  The phases for 3D carriers whose orbits are at maximal Fermi-surface cross-sections are also marked.  For the low-frequency carriers (Fig.~\ref{Flo:phase_summary_LF}), there is reasonable consensus that the carriers are SFs, with some disagreement over their dimensionality, probably because different types of graphite (natural, pyrolytic and HOPG) were used in different studies, and these had different inter-layer coupling.  However, for the high-frequency carriers (Fig.~\ref{Flo:phase_summary_HF}), there is no consensus with several authors, using different techniques, reporting phases inconsistent with either carrier nature. It is clear that there remains a controversy over the nature (DF or SF) of carriers in graphite, and also some concerns over the methods used to determine these from the phase of dHvA and SdH oscillations.  To resolve this, it is necessary to apply all the experimental techniques and analyses to the same sample.  In this paper we present the results of dHvA, SdH and Hall experiments carried out on the same sample of HOPG, covering a broader range of magnetic fields and temperatures than previous studies.

\section{Experimental Details\label{sub:experimental_details}}
\begin{figure}
\includegraphics[width=0.48\textwidth]{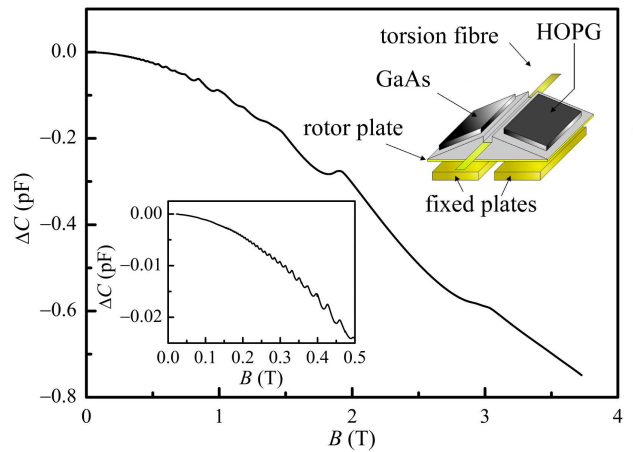}
\caption{Raw magnetometry data (change in magnetometer capacitance \textit{vs.} magnetic field) measured at 30$\,$mK.  The lower left inset shows the detail at low magnetic fields.  The upper inset is a schematic picture of the millikelvin torsion-balance magnetometer.  The sample magnetisation causes a twist of the rotor which results in an imbalance in the differential capacitor formed by the fixed and rotor plates.}
\label{Fig:raw_30mK}
\end{figure}

The sample used for the experiments detailed in this paper was a piece of ZYB grade HOPG measuring $10\times5.5\times0.5\,$mm. Magnetization measurements were made using a torsion-balance magnetometer shown schematically in the inset to Fig.~\ref{Fig:raw_30mK} and described in detail by \citet{Matthews2004}  The graphite sample had the normal to its atomic planes tilted, in two separate experiments, at 2$^\circ$ and at 20$^\circ$ to the applied magnetic field.  The magnetometer rotor was balanced using a piece of semi-insulating GaAs of similar mass.  Subsequently, the Hall and SdH measurements were performed on the same sample, contacted using silver paint, in six-terminal Hall-bar geometry. Standard low-frequency AC techniques were used with currents below 10$\,$\textmu A.  Measurements were carried out at temperatures in the range 30$\,$mK to 3.8$\,$K in the mixing chamber of a low-vibration dilution refrigerator.  The superconducting magnet used was found to have a remanent field of 26$\,$mT, which has been subtracted from all the results presented.  This remanent field was determined by analysing the (well-understood) dHvA oscillations in a GaAs heterojunction, and confirmed using a Hall-probe measurement.  

\section{Results}

\subsection{Hall Effect\label{sect:Hall}}

\begin{figure}
\includegraphics[width=\figwidth]{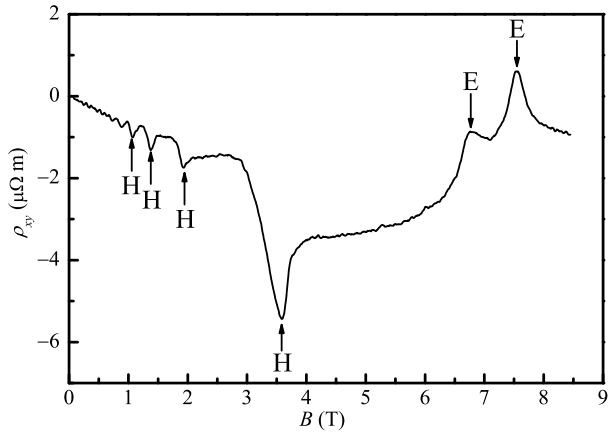}
\caption{Hall resistivity $\rho_{xy}$ plotted as a function of magnetic
field at 30$\,$mK .  Downward pointing features (marked H) occur when the Fermi energy passes through hole Landau levels; upward ones (E) occur when the Fermi energy passes through electron Landau levels.}
\label{Fig:rho_xy}
\end{figure}

Figure~\ref{Fig:rho_xy} shows the Hall resistivity $\rho_{xy}$ as a function of magnetic field at 30$\,$mK.  A large $\rho_{xx}$ signal, caused by imperfect contact geometry, was removed by subtracting the Hall resistances measured in forward and reverse magnetic fields. To understand the behaviour of the non-oscillating part of the Hall resistivity, we apply Drude theory, writing the conductivity as a sum of electron and hole contributions
\begin{eqnarray}
\label{Drude}
\sigma_{xx} &=&  \frac{n_e e \mu_e}{1 + \mu_e^2 B^2} + \frac{n_h e
\mu_h}{1 + \mu_h^2 B^2}
\ , 
\\
\sigma_{xy} &=&  - \frac{n_e e \mu_e^2 B}{1 + \mu_e^2 B^2} + \frac{n_h e
\mu_h^2 B}{1 + \mu_h^2 B^2}
\ ,
\end{eqnarray}
where~$n_e$, $n_h$ and~$\mu_e$, $\mu_h$ are carrier densities and mobilities. In a non-compensated system ($n_e \neq n_h$), the Hall resistivity in strong fields ($\mu_{e,h}B\gg1$) is determined by carrier density imbalance, $\rho_{xy} = B / e (n_h - n_e)$. In graphite, however, the carrier densities are compensated, $n_e \approx n_h$, and the Hall resistivity is due to non-equal mobilities, $\mu_e \neq \mu_h$. Assuming perfect compensation, $n_e = n_h = n$, 
we obtain  the Hall resistivity
\begin{equation}
\rho_{xy} = \frac{\sigma_{xy}}{\sigma_{xx}^2 + \sigma_{xy}^2} 
=  \frac{B}{en} \frac{\mu_h - \mu_e}{\mu_h + \mu_e}
\ .
\label{eq:rho_xy}
\end{equation}
(Interestingly, one can show that this relation holds at arbitrarily strong non-quantizing magnetic field.) Equation \ref{eq:rho_xy} shows that the negative slope of the Hall resistivity in Fig.~\ref{Fig:rho_xy} is due to lower hole mobility. Using the carrier density~$n = 3\times10^{18}\,{\rm cm}^{-3}$ obtained by~\citet{McClure1958} and the average slope of~$\rho_{xy}/B \approx 0.7 {\,\rm \mu \Omega \, m\,T}^{-1}$, we find the relationship between electron and hole mobilities to be~$\mu_e \approx 2 \mu_h$. 

The oscillations in $\rho_{xy}$ may be explained as follows. When the Fermi energy passes through a hole(electron) Landau level, scattering is enhanced and $\mu_{h}$($\mu_{e}$) is reduced causing a downward(upward)-pointing feature in $\rho_{xy}$.  Thus all the dips below 4$\,$T are due to hole Landau levels, while the twin peaks near 7.5$\,$T are caused by a spin-split electron Landau level.  This interpretation is in agreement with that of \citeauthor{Woollam1971} and enables us to confirm that the carriers with the higher oscillation frequency (higher value of $B_0$) are electrons while those with the lower frequency are holes (contrary to the assertion of \citeauthor{Lukyanchuk2004}).

\subsection{De Haas--van Alphen Effect\label{sub:dHvA}}

\begin{figure}
\includegraphics[width=\figwidth]{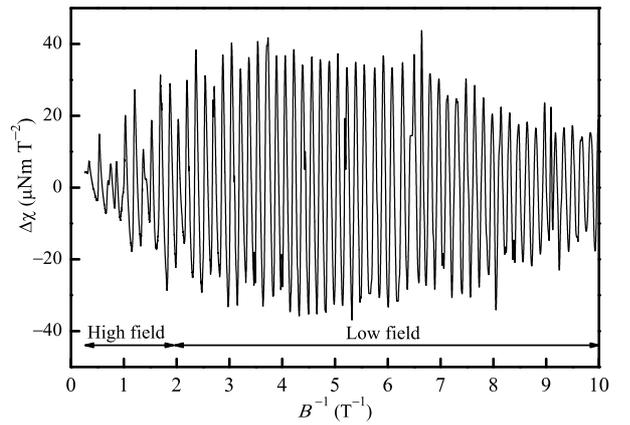}
\caption{Oscillatory magnetic susceptibility $\Delta\chi$ as a function of inverse magnetic field at
30$\,$mK. The arrows indicate the ranges of $B^{-1}$ used to obtain the FFTs of Fig.~\ref{Fig:30mK_dHvA_FFTs}.}
\label{Fig:delta_chi_30mK}
\end{figure}

A typical plot of magnetic susceptibility \textit{vs} $1/B$, taken at 30$\,$mK is shown in Fig.~\ref{Fig:delta_chi_30mK}.  The magnetic field in this and subsequent figures is the component of the applied field perpendicular to the graphite planes.  The magnetometer was calibrated \textit{in situ} using the known electrostatic force between the capacitor plates,\cite{Usher2009} and the raw capacitance data (Fig.~\ref{Fig:raw_30mK}) converted to torque and hence to magnetic moment, using $\mathbf{\tau}=\mathbf{m}\times\mathbf{B}$.  The magnetic moment is converted to magnetisation and then differentiated to give magnetic susceptibility.  The data in the figure have been filtered to remove high-frequency noise.  However, all the analysis presented in this paper has been carried out on unfiltered data.  The fast Fourier transform (FFT) algorithm used later requires that the susceptibility \textit{vs} $1/B$ data be interpolated to an integer power of 2, in this case $2^{14}$ points.

\subsubsection{Magnetic-Field and Temperature Dependences}
\label{sect:dHvA_B_and_T}

\begin{figure}
\includegraphics[width=\figwidth]{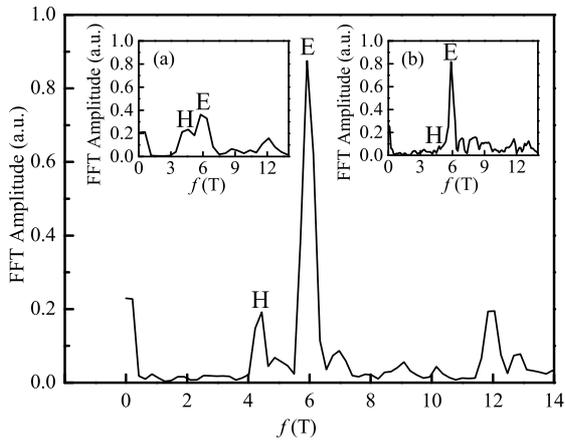}
\caption{FFT of the 30$\,$mK magnetic susceptibility (Fig.~\ref{Fig:delta_chi_30mK}) using data in the range from 0.27$\,$T$^{-1}$ to 5$\,$T$^{-1}$, showing the presence of hole oscillations with a fundamental field of \LB$\,$T, marked `H', and electron oscillations of approximately \HB$\,$T, marked `E'. The second harmonics of the two carriers are also visible at twice their fundamental frequencies. To illustrate the magnetic field dependences of the two sets of oscillations, the inset (a) shows the FFT from 0.27$\,$T$^{-1}$ to 2$\,$T$^{-1}$ (`High field' on Fig.~\ref{Fig:delta_chi_30mK}) where hole and electron peaks have roughly equal amplitude, while inset (b) shows the FFT for the data range from 2$\,$T$^{-1}$ to 10$\,$T$^{-1}$ (`Low field' on Fig.~\ref{Fig:delta_chi_30mK}) demonstrating the dominance of the electron oscillations at low $B$.}
\label{Fig:30mK_dHvA_FFTs}
\end{figure}

Figure~\ref{Fig:30mK_dHvA_FFTs} shows the FFT of the 30$\,$mK magnetic susceptibility data from 0.27$\,$T$^{-1}$ to 5$\,$T$^{-1}$.  Over this field range, both hole oscillations (FFT peak \LB$\,$T, marked H) and electron oscillations (\HB$\,$T, marked E) are visible, as well as their second harmonics. The insets show the FFT amplitudes for the high-field ((a), from 0.27$\,$T$^{-1}$ to 2$\,$T$^{-1}$) and low-field ((b), from 2$\,$T$^{-1}$ to 10$\,$T$^{-1}$) ranges respectively. In the high-field FFT both peaks are of roughly equal height, whereas in the low-field FFT, the electron peak dominates.  These data demonstrate that the (low-frequency) hole oscillation amplitude reduces more rapidly with decreasing field than that of the electrons. This implies either that the hole Landau levels are more closely spaced than the electron ones (the cyclotron masses satisfy the condition $m_{h}>m_{e}$), or that holes are more strongly scattered than electrons.
\begin{figure}
\includegraphics[width=\figwidth]{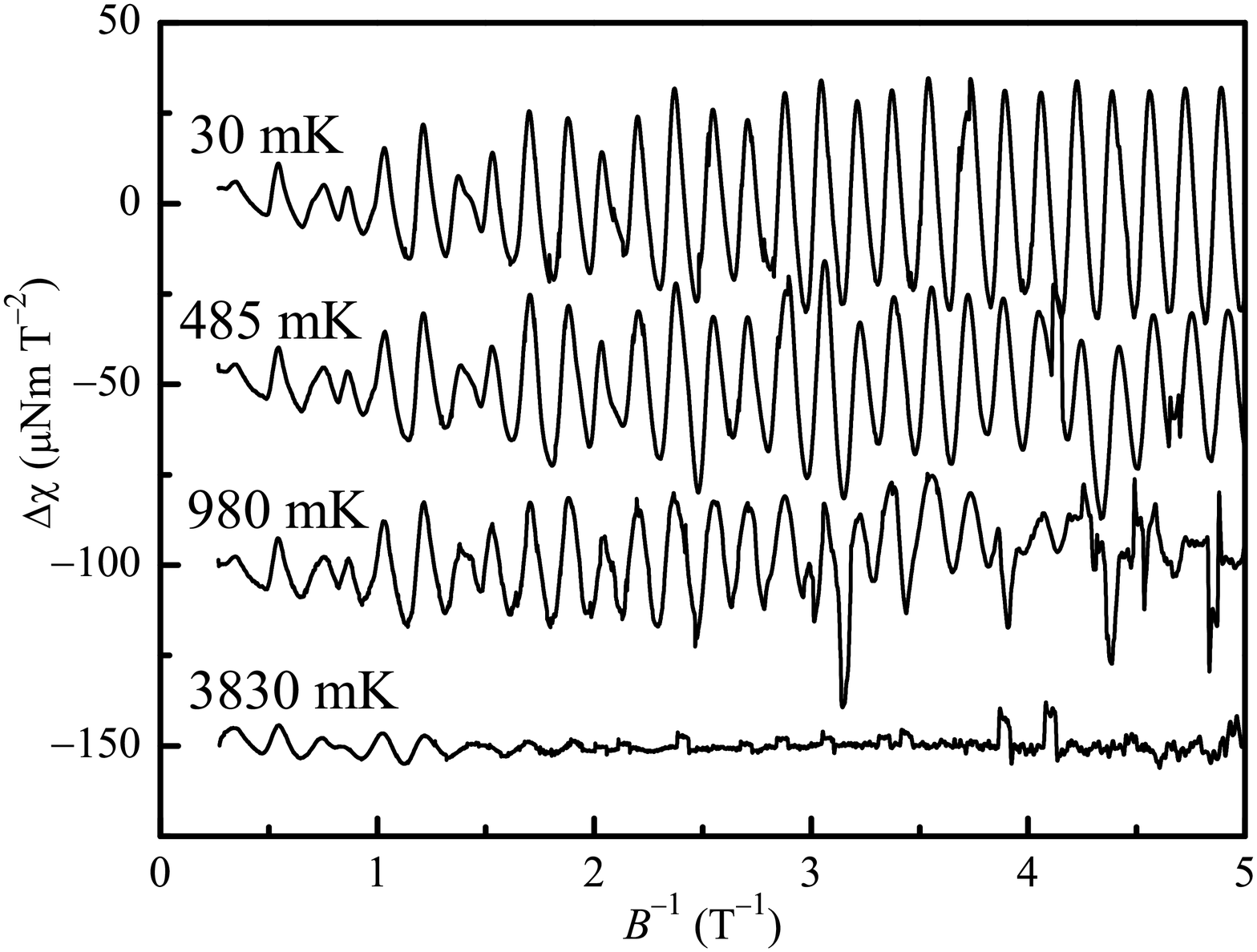}
\caption{dHvA oscillations at a range of temperatures. Oscillations at high $B^{-1}$ are predominantly due to electrons and these disappear rapidly as temperature increases leaving only the hole oscillations at low $B^{-1}$. Traces for successive temperatures are offset for clarity.}
\label{Fig:delta_chi_temperature_effects}
\end{figure}

\begin{figure}
\includegraphics[width=\figwidth]{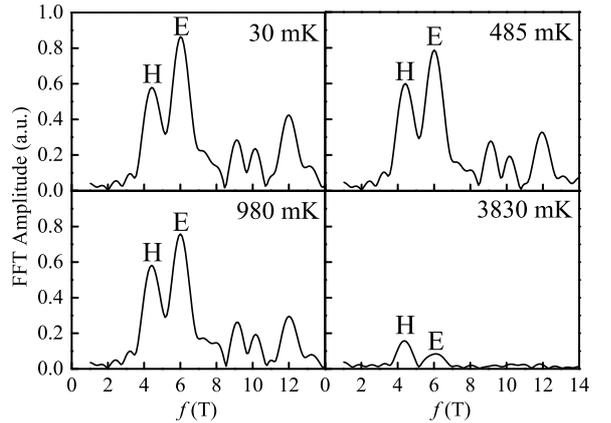}
\caption{FFT plots of the dHvA oscillations using data in the range from 0.27$\,$T$^{-1}$ to 2$\,$T$^{-1}$, for a range of temperatures. The high frequency (electron, E) peak is suppressed more rapidly with temperature than the low frequency (hole, H) peak.}
\label{Fig:FFT_temperature_effects}
\end{figure}

In order to distinguish between these two interpretations, we have examined the temperature dependence of the dHvA oscillations in Fig.~\ref{Fig:delta_chi_temperature_effects}. The oscillations at high $1/B$, which are predominantly due to electrons, diminish rapidly as the temperature increases, while the oscillations at low $1/B$ (holes) remain. This is shown explicitly in the FFTs of these data, in Fig.~\ref{Fig:FFT_temperature_effects}.  The fact that hole oscillations are more robust to temperature than those of electrons rules out the suggestion above that the hole Landau levels are more closely spaced than those of the electrons, and indeed indicates that the opposite is the case, $m_{h}<m_{e}$.  The observation that the high frequency carriers have larger mass than the low frequency ones is in agreement with \citet{Soule1964} who found cyclotron masses of 0.039$\,m_0$ and 0.057$\,m_0$ for the low-frequency and high-frequency carriers, respectively (although they assigned the wrong carrier types to these, as discussed above).  Thus the only possible interpretation of Fig.~\ref{Fig:30mK_dHvA_FFTs} is that the holes experience stronger scattering.  This is consistent with our assertion from the Hall data (Section \ref{sect:Hall}) that $\mu_e>\mu_h$. 

Various effects could cause different scattering rates for holes and electrons.  The scattering rate is determined by the integral of the density of states over the Fermi surface and the different densities of states for electrons and holes, caused by the different shapes of the Fermi surface at the $K$ point and near the $H$ points, will give different rates. The different natures of the carriers, DF or SF, also affect scattering.  It is known that in graphene and carbon nanotubes the chiral nature of the DFs suppresses back-scattering.\cite{Ando1998}  Further experiments on different types of graphite would help to clarify the causes of this effect.

\begin{figure}[ht!]
\centering
\subfloat[]{\includegraphics[width=\figwidth]{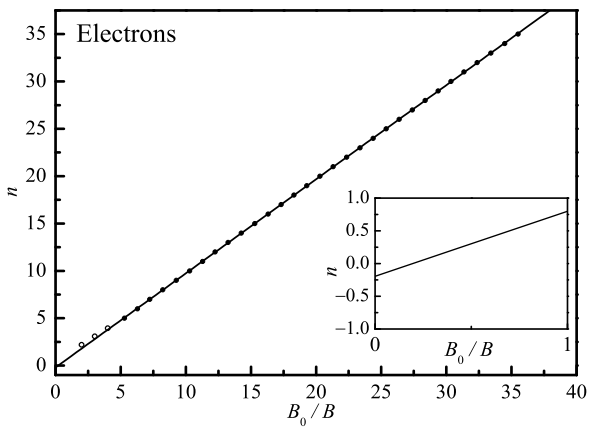}
\label{Flo:30mk_electrons_intercept_plot}}\\
\subfloat[]{\hspace{1pt}\includegraphics[width=0.437\textwidth]{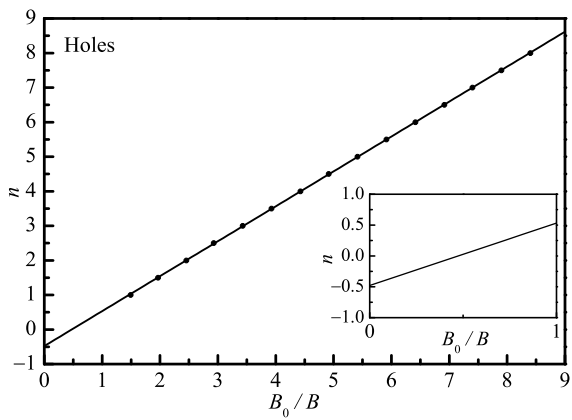}
\label{Flo:30mk_holes_intercept_plot}}

\caption{Intercept phase analysis of the 30$\,$mK dHvA oscillations. (a) Analysis of the oscillations of electrons. Peak position \textit{vs} Landau-level index $n$ is plotted up to $5\,$T$^{-1}$.  The intercept (see inset) indicates a phase of \Hphase$\,\left[2\pi\right]$. The open circles show how the low-index raw data deviate from the straight line.  (b) Analysis of the hole oscillation extrema (peaks and troughs) indicating a phase of \Lphase$\,\left[2\pi\right]$, consistent with 2D SFs.}
\label{Fig:30mK_intercept_plots}
\end{figure}

We can obtain quantitative information about the carrier effective masses and about scattering, from an analysis of the damping of the oscillations in magnetisation with magnetic field and temperature.  The amplitude of the magnetisation oscillations is given by:
\begin{equation}
\Delta M \propto \frac{\chi}{\sinh\chi} \exp\left(-\frac{\pi}{\omega_c\tau_q}\right)\ ,
\label{eq:envelope}
\end{equation}
where $\chi=2\pi^2k_BT/\hbar\omega_c$, $\omega_c=eB/m^*$ and $\tau_q$ is the quantum lifetime.  Determining $\tau_q$ is only possible for the carrier whose oscillations are dominant, in this case electrons.  We find that, for electrons, $m^*=0.046\pm0.003\,m_0$ and $\tau_q=1.7\pm0.1\times10^{-12}\,$s.  This effective mass is somewhat lower than the value reported by \citet{Soule1964} If we assume that $\tau_q$ is the same as the momentum relaxation time (the assumption of short-range scattering) then we obtain a mobility $\mu_e=6.7\pm0.4\,$m$^2$(V$\,$s)$^{-1}$.

\subsubsection{Carrier Natures -- Phase Analysis}

From the peaks in the FFT amplitudes the fundamental fields for the charge carriers have been identified. They are: low frequency holes  at \LB\LBerror$\,$T and high frequency electrons  at \HB\HBerror$\,$T. Following the approach of \citet{Lukyanchuk2004} and of \citeauthor{Schneider2009},\cite{Schneider2009} we can also extract the phase of the oscillations to determine the nature of the two types of carrier -- whether they are DFs or SFs. Two types of analysis can be applied: a plot of the $1/B$ positions of the oscillation extrema \textit{vs} Landau-level index yields an intercept from which the phase can be extracted; and the phase can be obtained directly from the FFT.

The phase is to be obtained using Eq. \eqref{eq:delta_chi_form} which applies to the fundamental frequency of the susceptibility oscillations. Since the raw data are not sinusoidal, they need to be filtered in order to perform an intercept analysis.  Fourier filtering was used, with a pass band centred approximately on the peak frequencies obtained from the FFT, and a band width sufficiently narrow to exclude the other peaks in the FFT.  Various combinations of band width and centre were tested, as well as other filter methods, and the intercept results were found to be fairly insensitive to these choices.

Figure~\ref{Fig:30mK_intercept_plots} shows the results of intercept phase analysis for the 30$\,$mK data of Fig.~\ref{Fig:delta_chi_30mK}. Low-index oscillation extrema (up to $n=4$) in the raw data  (open circles) are shown in the electron plot (a) to illustrate the deviation which has been commented upon by \citet{Schneider2009} and by \citeauthor{Smrcka2009},\cite{Smrcka2009} but are not used in the fit.  No such deviation is apparent in the hole data.  An intercept of zero corresponds to a total phase of zero (2D DFs from Table~\ref{Flo:carrier_nature_table}), while an intercept of $-0.5$ corresponds to a total phase of $-\pi$ (2D SFs).  The results are summarised in Table \ref{Flo:conclusion_table}.  From Fig.~\ref{Fig:30mK_intercept_plots}, the intercepts (in units of $2\pi$) are: for electrons, \Hphase, not consistent with any of the assignments in Table~\ref{Flo:carrier_nature_table}; for holes, \Lphase, consistent with their assignment as 2D SFs.  The error bars quoted for these intercepts reflect both the scatter of the straight-line graphs and small changes in intercept dependent on filtering parameters. 

\setlength{\tabcolsep}{2ex}
\begin{table*}
\begin{centering}
\begin{ruledtabular}
\begin{tabular}{c c cc cc c}
Carrier & Fundamental Field & \multicolumn{2}{c}{Total Phase from $\Delta\chi$} & \multicolumn{2}{c}{Total Phase from $\Delta\sigma_{xx}$} & Outcome \\
 & (T) & \multicolumn{2}{c}{$\left(\delta-\gamma\right)\,\left[2\pi\right]$} & \multicolumn{2}{c}{$\left(\delta-\gamma\right)\,\left[2\pi\right]$} & \\
 & & Intercept & FFT & Intercept & FFT & \\
\hline\\[-5pt]
Electrons 	& \HB \HBerror 	& \Hphase 	& \HphaseFFT 	& - 			& - 				& indeterminate \\
Holes 	& \LB \LBerror 	& \Lphase  	& \LphaseFFT 	& \SdHphase 	& \SdHphaseFFT 	& 2D SF \\
\end{tabular}
\end{ruledtabular}
\par\end{centering}
\caption{Experiment conclusions for the nature of carriers in graphite.}
\label{Flo:conclusion_table}
\end{table*}

\begin{figure}
\includegraphics[width=\figwidth]{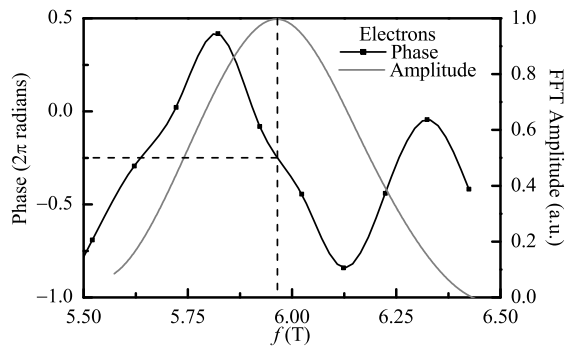}
\caption{FFT phase analysis of the 30$\,$mK dHvA results for electrons - a phase of \HphaseFFT$\,\left[2\pi\right]$ is observed. The error arises from the uncertainty in determining the peak position.}
\label{Fig:30mK_phase_plot}
\end{figure}
\begin{figure}[ht!]
\subfloat[]{\includegraphics[width=\figwidth]{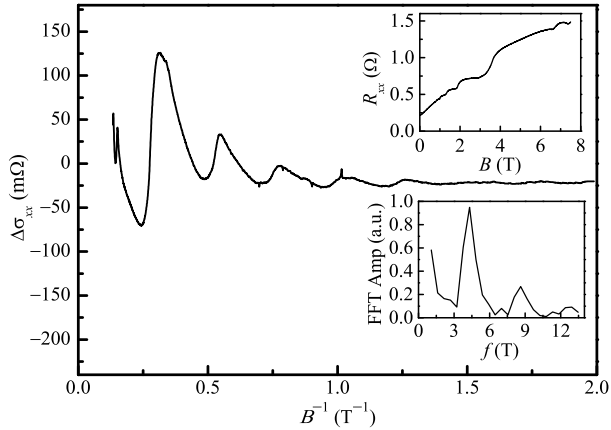}
\label{Flo:30mk_delta_sigma_xx}}\\
\subfloat[]{\includegraphics[width=0.43\textwidth]{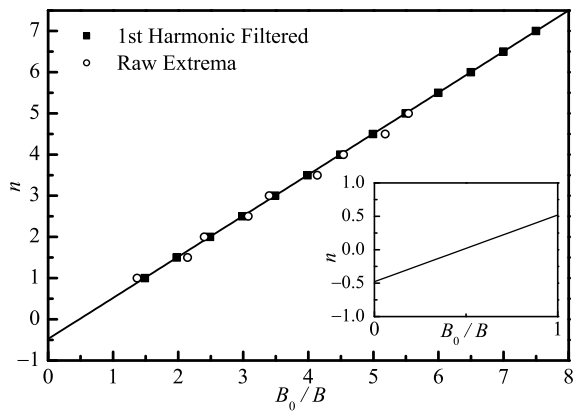}
\label{Flo:30mk_holes_intercept_transport}}

\caption{(a) $\Delta\sigma_{xx}$ as a function of inverse magnetic field at 30$\,$mK. The upper inset shows the raw $R_{xx}$ data as a function of magnetic field.  The lower inset is the FFT of $\Delta\sigma_{xx}$; the low frequency (hole) peak and its first harmonic are visible but the high frequency (electron) peak is not. (b) Intercept phase analysis of the SdH oscillation extrema (peaks and troughs).  Only holes were could be analysed and first-harmonic filtered data were used as discussed in the text, yielding a phase of \SdHphase$\,\left[2\pi\right]$, consistent with 2D SF carriers.  Raw extrema are also shown but exhibit systematic shifts to the right and to the left of the filtered data due to the presence of higher harmonics.}

\label{Fig:transport_30mK}
\end{figure}
Phase information was also extracted directly from the FFTs. The phase was found to vary rapidly with field close to the FFT peaks (Fig.~\ref{Fig:30mK_phase_plot}), and consequently the uncertainty in the phase was governed by the accuracy with which one could determine the peak positions, which in turn was controlled by the range in $1/B$ of the susceptibility data. With the data spacing of Fig.~\ref{Fig:30mK_phase_plot} we estimate the uncertainty in the position of the FFT peak to be about $\pm0.1\,$T leading to an error in the phase of about $\pm\pi/2$.  Consequently, this gives a reasonable indication of carrier nature, but is not sufficiently accurate to determine dimensionality.  Nevertheless the results of this method are in very good agreement with those of the intercept method (Table \ref{Flo:conclusion_table}): the phases (in units of $2\pi$) are: for electrons, \HphaseFFT, the error bar being too large to determine the nature of the carriers; for holes, \LphaseFFT, consistent with their assignment as SFs.

\subsection{Shubnikov--de Haas Effect}

Figure~\ref{Flo:30mk_delta_sigma_xx} shows the oscillating part of $\sigma_{xx}$, $\Delta\sigma_{xx}$, as a function of $1/B$ at a temperature of 30$\,$mK.  It was extracted from the raw longitudinal resistance $R_{xx}$ data (upper inset to Fig.~\ref{Flo:30mk_delta_sigma_xx}) using
\begin{equation}
\Delta\sigma_{xx}=\Delta\frac{\rho_{xx}}{\rho_{xx}^{2}+\rho_{xy}^{2}}\simeq-\frac{\Delta\rho_{xx}}{\rho_{xx}^{2}}\sim-\Delta R_{xx}.
\label{eq:delta_sigma_xx}
\end{equation}
The oscillations up to 4$\,$T correspond to those seen in the Hall effect, which have been identified as hole oscillations.  Only the two features around 7$\,$T are associated with electrons. The FFT of $\Delta\sigma_{xx}$ (lower inset to Fig.~\ref{Flo:30mk_delta_sigma_xx}) shows a clear peak due to hole oscillations but no evidence of electrons.  This is consistent with the lack of electron-related features in the raw data, but contrasts with the analysis of the susceptibility data.  

To understand the absence of electrons in the SdH data, we note the following. For a conductor with only one carrier type, the SdH oscillations and dHvA oscillations are both related to the oscillations in the density of states, and are therefore proportional to each other (see, e.g. Section 11.1 of \citet{Abrikosov1988}):
\begin{equation}
\frac{\delta \sigma_{xx}}{\sigma_{xx}} \sim 
\left(\frac{m^\ast B_0}{B} \right)^2
\frac{\partial M}{\partial B}
\ , 
\end{equation}
where~$m^\ast$ is the cyclotron mass, and~$B_0$ is the fundamental field. Note that the oscillating correction to the conductivity is proportional to the conductivity itself. The latter is inversely proportional to the mobility in strong fields:
$\sigma_{xx} \propto 1/(\mu B^2)$. Therefore, the amplitude of conductivity oscillations at high fields, when the Dingle factor can be neglected, is higher if the mobility is {\it lower}. This is because, while thermodynamic dHvA oscillations are entirely determined by the density of states, SdH oscillations are caused by the enhancement of the scattering rate by the peaks in the DOS. For this reason, they also include the scattering rate as a prefactor. 

In graphite, different carrier types contribute independently to conductivity oscillations. The relative magnitude of such contributions depends on the scattering rate for a given carrier type. Thus, high-mobility electrons should contribute {\it less} to the SdH oscillations than low-mobility holes.

Figure~\ref{Flo:30mk_holes_intercept_transport} shows the intercept phase analysis for the SdH data. The SdH data have significant harmonic content in the range over which the analysis is carried out as can be seen from their FFT (lower inset of Fig.~\ref{Flo:30mk_delta_sigma_xx}). The same filtering process as described for the dHvA data above was used to obtain extrema positions in Fig.~\ref{Flo:30mk_holes_intercept_transport}. The raw extrema are shown for comparison. A phase of \SdHphase$\,\left[2\pi\right]$ is in excellent agreement with the analysis of the dHvA data (Fig.~\ref{Flo:30mk_holes_intercept_transport}) once again identifying the holes as 2D SFs. FFT phase analysis of the SdH data yields a hole phase of \SdHphaseFFT$\,\left[2\pi\right]$. This is in agreement with the intercept phase and the phases from the dHvA analysis, and indicates that holes are SFs, but again is subject to a substantial error.

The amplitude $\Delta\rho_{xx}$ of the resistivity oscillations has the same magnetic field and temperature dependences as $\Delta M$ (Eq. \ref{eq:envelope}). Analysis of this amplitude gives, for holes, $m^*=0.033\pm0.003\,m_0$ and $\tau_q=2.1\pm0.2\times10^{-13}\,$s.  This effective mass is in reasonable agreement with the value reported by \citet{Soule1964} Assuming that $\tau_q$ is the same as the momentum relaxation time we obtain a hole mobility $\mu_h=1.2\pm0.1\,$m$^2/$Vs.  The dHvA amplitude calculation for electrons (Section \ref{sect:dHvA_B_and_T}) gave a mobility of $\mu_e=6.7\pm0.4\,$m$^2/$Vs.  The amplitude analyses therefore confirm qualitatively the conclusion from the Hall effect data (Section \ref{sect:Hall}) that $\mu_e >  \mu_h$. Quantitative agreement is not expected because the Hall effect is sensitive to relaxation times averaged over the entire Fermi surface while magneto-oscillatory effects primarily probe behaviour at the extremal cross-sections.

\section{Conclusions}

We have carried out Hall, SdH and dHvA measurements on the same sample of HOPG, at temperatures down to 30$\,$mK. Our Hall effect experiments confirm the carrier types established by \citet{Schroeder1968}: holes near the $H$ point of the graphite Brillouin zone are responsible for the low-frequency component of the dHvA oscillations, while electrons at $K$ give rise to the high-frequency component.  Analyses of the Hall, dHvA and SdH effects all indicate that the holes are subject to stronger scattering than the electrons.  The origin of this observation will be the subject of further investigations of various types and grades of graphite.  The temperature dependence of the dHvA effect indicates that the holes have larger Landau-level separations than the electrons, a conclusion confirmed by amplitude analyses of dHvA and SdH oscillations. 

The  analysis of the phase of the dHvA and SdH oscillations suggests that the low-frequency carriers (holes) are SFs, in agreement with the previous studies summarised in  Table~\ref{Flo:previous_conclusions_table} and Fig.~\ref{Fig:phase_summary}.  The phases are consistent with the holes being two-dimensional, in agreement with the HOPG experiments of \citet{Schneider2009} but not with other experiments.  This probably reflects the fact that the interaction between atomic layers is different in the different samples used.  The situation with the high-frequency carriers (electrons) is more complicated.  Our results show an intermediate value of the phase, consistent with neither DF nor SF.  Other authors (\citeauthor{Williamson1965},\cite{Williamson1965} \citet{Schneider2009}) have also reported an intermediate phases for the high-frequency carrier.  It is interesting to note that \citet{Mikitik2006} pointed out that intermediate phases are possible for minority holes in situations close to magnetic breakdown, though this could not occur for majority carriers with standard values of the SW parameters. Intermediate phases might also be the result of sample inhomogeneity, in which electrons are SF and DF in different parts of the same sample.  It is not clear why this should affect electrons more than holes though.  Clearly graphite remains an interesting and intriguing material and  more measurements on a range of graphite samples are needed in order to determine the nature of its electrons.

\bibliography{references} 

\end{document}